\documentclass[aps,prl,twocolumn,groupedaddress,showpacs]{revtex4}
\usepackage{epsfig}
\usepackage{epsf}
\usepackage{amssymb, amsmath}
\usepackage{graphicx}
\usepackage{dcolumn}
\usepackage{bm}
\usepackage{appendix}

\usepackage{color}
\usepackage{ulem}
\begin{document}
\title{Change of an insulator's topological properties by a Hubbard interaction}
\author{Miguel A. N. Ara\'ujo$^{1,2,3}$, Eduardo V. Castro$^{1,3}$, 
and Pedro D. Sacramento$^{1,3}$
} 
\affiliation{$^1$ CFIF, Instituto Superior 
T\'ecnico, TU Lisbon, Av. Rovisco Pais, 1049-001 Lisboa, Portugal}
\affiliation{$^2$ Departamento de F\'{\i}sica,  Universidade de \'Evora, P-7000-671, \'Evora, Portugal}
\affiliation{$^3$ Beijing Computational Science Research Center, Beijing 100089, China}

\begin{abstract}
We introduce two dimensional fermionic band models with two orbitals per lattice site, 
or one spinful orbital, and which 
have a non-zero topological Chern number that can be changed by varying the ratio of hopping  parameters. 
A topologically non-trivial insulator is then realized if there is one fermion per site. 
When interactions in the framework of the Hubbard model are introduced, the effective hopping parameters are 
renormalized and the system's topological number can change at a certain interaction strength, $U=\bar U$, 
smaller than that for the Mott transition. Two different situations may then occur: either 
the anomalous Hall conductivity $\sigma_{xy}$ changes abruptly at $\bar U$, 
as the system undergoes a transition from one topologically non-trivial insulator to another, 
or the transition is through an anomalous Hall metal, and $\sigma_{xy}$ changes smoothly between two 
different quantized values as $U$ grows. Restoring time-reversal symmetry by adding spin to spinless models, 
the half-filled system becomes a $\mathbb{Z}_2$ topological insulator. The topological number $\nu$ then changes 
at a critical coupling $\bar U$ and the quantized spin Hall response changes abruptly.
\end{abstract}

\maketitle


Recent interest in non-trivial topological properties of insulators \cite{hasankane,QiZhang} has spurred intensive  search for band  models with non-trivial topology. This is because of the possibility that electron  interactions in fractionally filled topologically non-trivial bands may lead to the realization of highly correlated fractional quantum Hall states \cite{dasSarma1,mudry,wen,Fito}.  On the other hand, cold atomic gases in  optical lattices with  tunable interaction strength open the possibility to physically realize topological insulators \cite{opticallattice,sun2}. It is then natural to ask about the effects that electron correlations can have on the topological properties for proposed models.
 
Indeed, non-interacting topological phases are fairly well understood, so a great deal of attention has recently been given to the effect of interactions 
\cite{balents,DungHaiLee,Varney_1,varney_2,Rossier,Culcer,yoshida_1,yoshida_2,medhi,castro,leHur,chineses,muramatsu,assad,ShunYu,raghu,JunWen,thomale,dauphin,sun,Uebelacker,weeks}.
Studies of the Kane-Mele-Hubbard model \cite{haldane1988,KaneMele} have recently been carried out, showing that the topological insulator survives until the Mott insulating phase is attained 
\cite{leHur,chineses,muramatsu,assad,ShunYu}. 
A  topological phase can also arise from interactions added to a trivial band model, 
leading to a topological Mott insulator \cite{raghu,JunWen,thomale,dauphin,weeks,Wang,Ma}. Also,
the  instability of quadratic band crossings to arbitrarily weak interactions
has been proved \cite{sun,Uebelacker,weeks}.

 In the present paper we show that yet another possibility exists. Namely, that a {\it purely local} 
interaction, such as that in the  Hubbard model,  
can drive the system from one topologically non-trivial insulating phase into another by changing its topological number, while keeping it finite. This mechanism is particularly relevant for proposed band models with Chern number $C$ larger than one.
We present band models which break time-reversal symmetry explicitly, a situation  analogous to the one considered by Haldane for the honeycomb lattice \cite{haldane1988}. The system is a quantum anomalous Hall insulator when the lowest band is filled. If a Hubbard repulsive interaction is also present, the topological Chern number $C$, hence the number of chiral edge states, changes at critical values of the Hubbard interaction strength, $\bar U$,  before a Mott insulator phase is attained at higher $ U_c > \bar U$. Such transition is signaled  by a  change in the quantized  Hall response. This effect occurs because the interaction effectively renormalizes the Hamiltonian parameters for the fermions, decreasing the longer ranged hopping with respect to the short ranged. Such changes in the  effective hopping parameters induce a change of the topological number of bands for the fermions. 
 
The following example models describe  electrons in a square lattice with  two orbitals per site. 
 The Pauli matrices
  $\tau_\mu$ and $\sigma_\mu$ ($\mu=0,1,2,3$) act on the orbital (or sub-lattice) space and
 spin space, respectively,  and the subscript ``0'' refers to the identity matrix.
 The  Hamiltonian has the general form:
 \begin{equation}
\hat H(\boldsymbol h) = \boldsymbol h( \boldsymbol k) \cdot  \boldsymbol \tau
  + h_0(\boldsymbol k) \tau_0 \,,
\label{H}
\end{equation}
where $ \boldsymbol h=(h_x,h_y,h_z)$ 
and $\boldsymbol k=(k_x,k_y)$ denotes the momentum vector.  The 
Chern number for  the bands in  Hamiltonian Eq.~(\ref{H})  is independent of the choice for 
$h_0(\boldsymbol k)$, as computed from the
usual expression 
\begin{equation}
C=\frac{1}{4\pi}\int dk_x\ dk_y \  \frac{\partial \hat{\boldsymbol h}}{\partial k_x} 
\times  \frac{\partial \hat{\boldsymbol h}}{\partial k_y} \cdot \hat{\boldsymbol h}\,,
\label{chern}
\end{equation}
and we therefore neglect $ h_0(\boldsymbol k) $ for the time being, and comment on it later.
Time-reversal symmetry (TRS) requires $h_{x(z)}$ to be  a even function 
of $\boldsymbol k$ and $h_y$ to be  odd. 
In order to have nonzero $C$, TRS must be broken. The operation
of spatial inversion [${\cal I}: h_x(\boldsymbol k) \longrightarrow h_x(-\boldsymbol k)$,
$h_{y(z)}(\boldsymbol k) \longrightarrow -h_{y(z)}(-\boldsymbol k)]$
 does not change $C$.

\begin{figure}
\begin{centering}
\includegraphics[width=7cm]{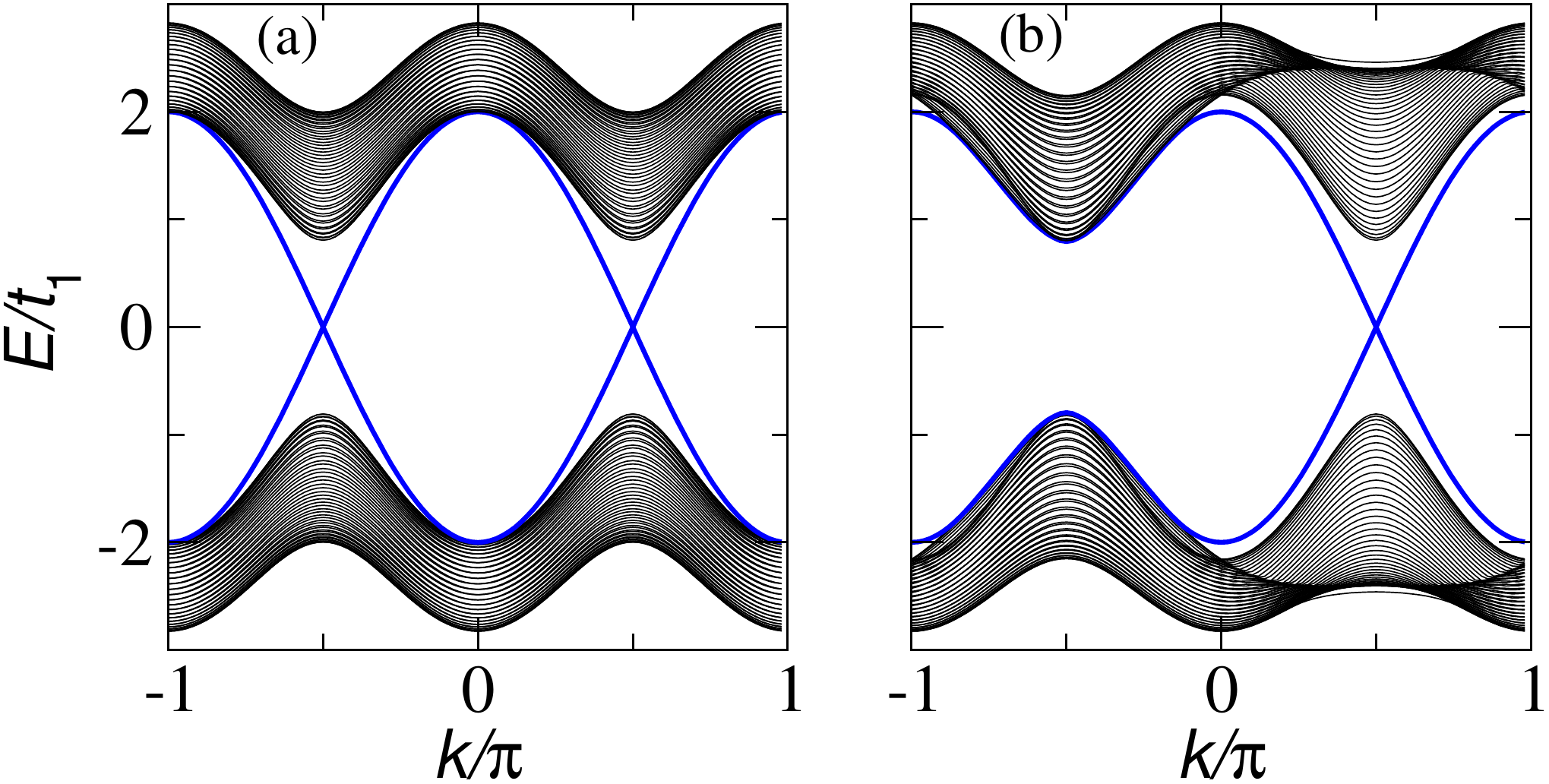}
\par\end{centering}
\caption{\label{fig:edges}(color online). Spectrum for the model given by 
Eq.~(\ref{model1}) in the ribbon geometry: (a) for $t_2 >t_1'-\delta/4$, where $C=2$; (b) for $t_2 <t_1'-\delta/4$, where $C=1$.}
\end{figure} 
%
  
{\it Model for spinless fermions with $\mathbb{Z}$ topological number.---}We 
consider
\begin{eqnarray}
h_x &=&
   \sqrt{2} t_1  \left( \cos k_x + \cos k_y \right)\,,
\nonumber\\
h_y 
&=&  \sqrt{2} t_1  \left( \cos k_x - \cos k_y \right)   \,,  \label{model1}\\
h_z &=& 4t_2 \sin k_x \sin k_y + 2 t_1' \left(  \sin k_x + \sin k_y  \right)  + \delta
\nonumber
 \,.
\end{eqnarray} 
The terms $t_1'$ and $h_y$  break TRS and
are responsible for a non-zero Chern number. The terms  $t_2$ and $h_y$ break
spatial  inversion symmetry.
 If  $t_2 >t_1'-\delta/4$ then Eq.~(\ref{chern}) gives  $C=2$, and if 
 $t_2 < t_1' - \delta/4$ then  $C=1$.
The spectrum obtained in the ribbon geometry is shown in Fig.~\ref{fig:edges}(a) for $t_2 >t_1'-\delta/4$, the $C = 2$ case. Two pairs of counter propagating edge modes, running along the opposite edges of the ribbon, are clearly seen. The case $t_2 <t_1'-\delta/4$ is shown in  Fig.~\ref{fig:edges}(b). Compatible with $C=1$, now only one edge mode runs along each edge of the ribbon.
We take the case where the lowest band is filled with spin polarized (or spinless)
electrons, that is, where  one  electron occupies one of the  two orbitals per lattice site. 
 The system is then a band insulator with a number of chiral
edge modes equal to the Chern number (see Fig.~\ref{fig:edges}).
We now assume that the parameters  in Eq.~(\ref{model1})
are such that $C= 2$ and introduce electron-electron
interactions via the Hubbard term:
\begin{equation}
\hat H_{int}=  \frac{U}{2} \sum_j \left( \sum_{s}   \hat n_{j,s}   - 1   \right)^2
 \end{equation}
where $s=1,2$ denotes the orbital index  at lattice site $j$
  and $ \hat n_{j,s}$ is the electron number operator.
A suitable approach to the Hubbard model at half-filling is 
the slave-rotor method \cite{FlorensGeorges,FlorensGeorges2,SSLee}.
In this approach a rotor $e^{i\theta_j}$ is assigned to every  lattice site and 
the charge at every site  is identified with the rotor's angular momentum, 
$\hat L_j =  \sum_{s}   \hat n_{j,s}   - 1  =  -i\partial /\partial\theta_j$.
The electron operator
is decomposed into a fermion and a rotor as 
$\hat c_{j} = \hat f_{j} e^{-i\theta_j}\,.$
The interaction can then be simply rewritten as:
$
\hat H_{int}=  \frac{U}{2} \sum_j \hat L_j ^2\,.
$
We shall employ the $X$-boson treatment \cite{FlorensGeorges}  by equating $e^{i\theta_j}=X_j$ 
with the constraint  $|X_j|^2=1$. 
Such a treatment has the advantage that it allows  the calculation of 
the rotor correlation function between arbitrary sites of the lattice, as the
decay of this correlation function with distance in the bose condensed phase is of
 crucial importance in what follows.  Previous studies with this method
 concentrated on nearest  neighbor boson correlation and were focused on the
 Mott transition \cite{leHur,balents}.
The method is well explained in Ref.~\cite{FlorensGeorges} 
and has been applied to the study of Mott transitions in topological 
insulators a number of times.
The Hamiltonian Eq.~({\ref{H}}) must be Fourier transformed
  to real space in order to  make the substitution 
$\hat c_{j} = \hat f_{j} e^{-i\theta_j}\,$
and becomes 
a function of $f$-fermions  and $X$-bosons, $\hat H(\hat f^\dagger X,\hat  f X^*)$. 
 The partition function is ${\cal Z} = \int D\lambda D\bar f Df DX^* DX \exp[-S]$ 
where  the action $S$ reads
\begin{eqnarray}
S= \int_0^{1/T}d\tau \left[   \sum_{j,s}
\bar f_{j,s} (\partial_\tau - \mu) f_{j,s} + \sum_j  \frac{1}{2U}|\partial_\tau X_j |^2
\right. \nonumber\\
  + H(\bar f X,  f X^*) 
+ \sum_j  i\lambda_j \left(  |X_j |^2  - 1   \left.  \right)
\right]
\end{eqnarray}
The imaginary time $\tau$ is not to be confused with the Pauli matrices above. 
The Hamiltonian in the action reads
\begin{eqnarray}
 H(\bar f X,  f X^*) = \sum_{ij}\sum_{ss'}
{\boldsymbol h}_{ij}\cdot {\boldsymbol \tau}^{ss'}  \bar f_{i,s}  f_{j,s'}  X_{i}X_{j }^*\,,
\label{hfx}
\end{eqnarray}
with
${\boldsymbol h}_{ij} = \frac{1}{N_s} \sum_{\boldsymbol k} 
{\boldsymbol h}({\boldsymbol k})
e^{i{\boldsymbol k}\cdot ({\boldsymbol r}_i-{\boldsymbol r}_j)}$,
where $N_s$ denotes the number of lattice sites.
We  do   the standard mean-field decoupling of the fermions 
and bosons \cite{FlorensGeorges,leHur}, whereby 
the Hamiltonian becomes a sum of a fermionic and a bosonic term, 
$H(\bar f X,  f X^*)= H_f(\bar f,f) + H_X(X,X^*)$.
 The $f$-fermion Hamiltonian,  $\hat H_f$ back in momentum space 
  is given by expressions in Eq.~(\ref{H}) 
 and Eq.~(\ref{model1}) but where the hopping
 parameter between sites $i$ and $j$  is now multiplied by $\langle  X_i X_j^* \rangle$, 
{\it i.e.}, 
 $ {\boldsymbol h}_{ij} \rightarrow {\boldsymbol h}_{ij}\langle  X_i X_j^* \rangle$.
The boson Hamiltonian has the form:
\begin{eqnarray}
H_X &=&  \sum_{ij} {\boldsymbol h}_{ij}\cdot \langle \sum_{ss'}{\boldsymbol \tau}^{ss'}  \bar f_{i,s}  f_{j,s'} \rangle  X_{i}X_{j }^*\nonumber\\
&=& J_1 \sum_i\sum_{j'} 
 X_iX_j^* + J_2  
 \sum_i\sum_{j''}
 X_iX_j^*
\end{eqnarray}
where  $j'$, $j''$ denote first and second neighbors to the site $i$, respectively, and
where $J_{1(2)}$ is obtained by  averaging the fermion fields in Eq.~(\ref{hfx}). 
$H_X$, enjoys in this case  the square lattice symmetry.
The action is, accordingly, a sum of two parts, $S=S_f+S_X$, with
\begin{eqnarray*}
S_X = \int_0^{1/T}d\tau \left[  \sum_j
\frac{1}{U}|\partial_\tau X_j |^2 +  \sum_j  i\lambda_j \left(  |X_j |^2  - 1  \right)
+ H_X 
\right]\,,  
\end{eqnarray*}
where the rescaling $U \rightarrow U/2$ has been made in order for the slave-rotor method 
to reproduce the correct atomic limit \cite{FlorensGeorges,FlorensGeorges2}.
In momentum space and Matsubara frequency, the  boson field can be written as
\begin{eqnarray}
X_j(\tau) = \frac{1}{\sqrt N_s}  \sum_{i\nu, \boldsymbol k}  
X(i\nu, \boldsymbol k) e^{i(\boldsymbol k\cdot {\boldsymbol r}_j - \nu \tau)} + \sqrt{x_0}
\label{Xdecomp}
\end{eqnarray}
where $x_0$ is a density of the condensate of the boson field at zero momentum and frequency. 
The summation in Eq.~(\ref{Xdecomp}) excludes 
the point $(i\nu, \boldsymbol k)=(0, \boldsymbol 0)$.
The action for the bosons then reads 
\begin{eqnarray}
S_X = \frac{1}{T}  \sum_{i\nu, \boldsymbol k}  |X(i\nu, \boldsymbol k)|^2
 \left[
 \frac{\nu^2}{U} + 
J_1 \gamma_1( {\boldsymbol k} ) + J_2 \gamma_2({\boldsymbol k} ) + i\lambda
\right]\,.
\nonumber\\
\end{eqnarray}
with $ \gamma_1( {\boldsymbol k} ) = 2\cos(k_x) + 2\cos(k_y)$ and  $ \gamma_2( {\boldsymbol k} ) = 4\cos(k_x) \cos(k_y)$.
Assuming a spatially constant $i\lambda_j $, the constraint $|X_j|^2=1$ can be implemented on average in imaginary time
and space as
\begin{eqnarray}
1=x_0 +
{\frac{\sqrt{U}}{2}} \frac{1}{N_s} \sum_{\boldsymbol k}
 \frac{1}{\sqrt{ J_1 \gamma_1( {\boldsymbol k} ) + J_2 \gamma_2({\boldsymbol k} ) + i\lambda }}
\end{eqnarray}
If the interaction is not strong enough to enter in the Mott regime,  
the condensate density is finite and this requires 
the dispersion relation for the bosons $\sqrt{ J_1 \gamma_1( {\boldsymbol k} ) + J_2 \gamma_2({\boldsymbol k} ) + i\lambda }$
to vanish at ${\boldsymbol k}=0$. The Lagrange multiplier is then fixed to the value
$i\lambda = -4(J_1 + J_2)$. The instantaneous boson spatial correlation function reads
\begin{eqnarray}
\langle X_i X_j^* \rangle =
1-
{\frac{\sqrt{U}}{2}} \frac{1}{N_s} \sum_{\boldsymbol k}
 \frac{1- e^{i\boldsymbol k\cdot ({\boldsymbol r}_j - {\boldsymbol r}_j) }}
  {\sqrt{ J_1 \gamma_1( {\boldsymbol k} ) + J_2 \gamma_2({\boldsymbol k} ) + i\lambda }}\,.\nonumber\\ 
  \label{XX}
\end{eqnarray}
The result in Eq.~(\ref{XX}) must be inserted back into the $f$-fermion Hamiltonian $H_f$, until convergence is attained.

A  simulation with 
$t_1=1$, $t_2=0.7$,  $t_1' = 0.8t_2$ and $\delta=0$, 
yields  a transition 
from the Chern number $C=2$ at $U<\bar U$   to $C=1$ at   $U>\bar U$  where $\bar U \approx 1.4$. The 
critical interaction for the Mott transition obtained is $U_c \approx 2.9$. 
For  $U>\bar U$  we then expect a discontinuous change in the  Hall conductivity. 
At $U=\bar U$ the $f$-fermion bands touch and close the gap.
As long as $t_2 >t_1'-\delta/4$, there is a $\bar U<U_c$. 
If   $t_2 <t_1'-\delta/4$, then $C=1$ for all $U<U_c$. 

On the Mott insulating side,  $U>U_c$,  the condensate $x_0$ vanishes 
and the boson correlation function is proportional to $\sqrt{U}$, so that 
the ratio between first and second neighbors, 
$ \langle  X_i X_{i+\hat x}^* \rangle/ \langle  X_i X_{i+\hat x+\hat y}^* \rangle$, 
is independent of $U$. 
Therefore, the above  mechanism is not effective inside the Mott phase.

The physical electron's Green's function has a coherent part where 
the  boson condensate $x_0$ is the quasi-particle weight and the excitations
are those of the Hamiltonian $H_f$.  We have checked that the Chern number
obtained from $H_f$ and Eq.~(\ref{chern})
agrees with  that obtained from the physical
electron's full Green's function for an interacting system (equation (6) in 
Ref.~\cite{wangzhang}).
In the Green's function point of view, the change in the topological number
occurs because a pole of the Green's function moves across  zero energy
\cite{gurarie}.

\begin{figure}
\begin{centering}
\includegraphics[width=7cm]{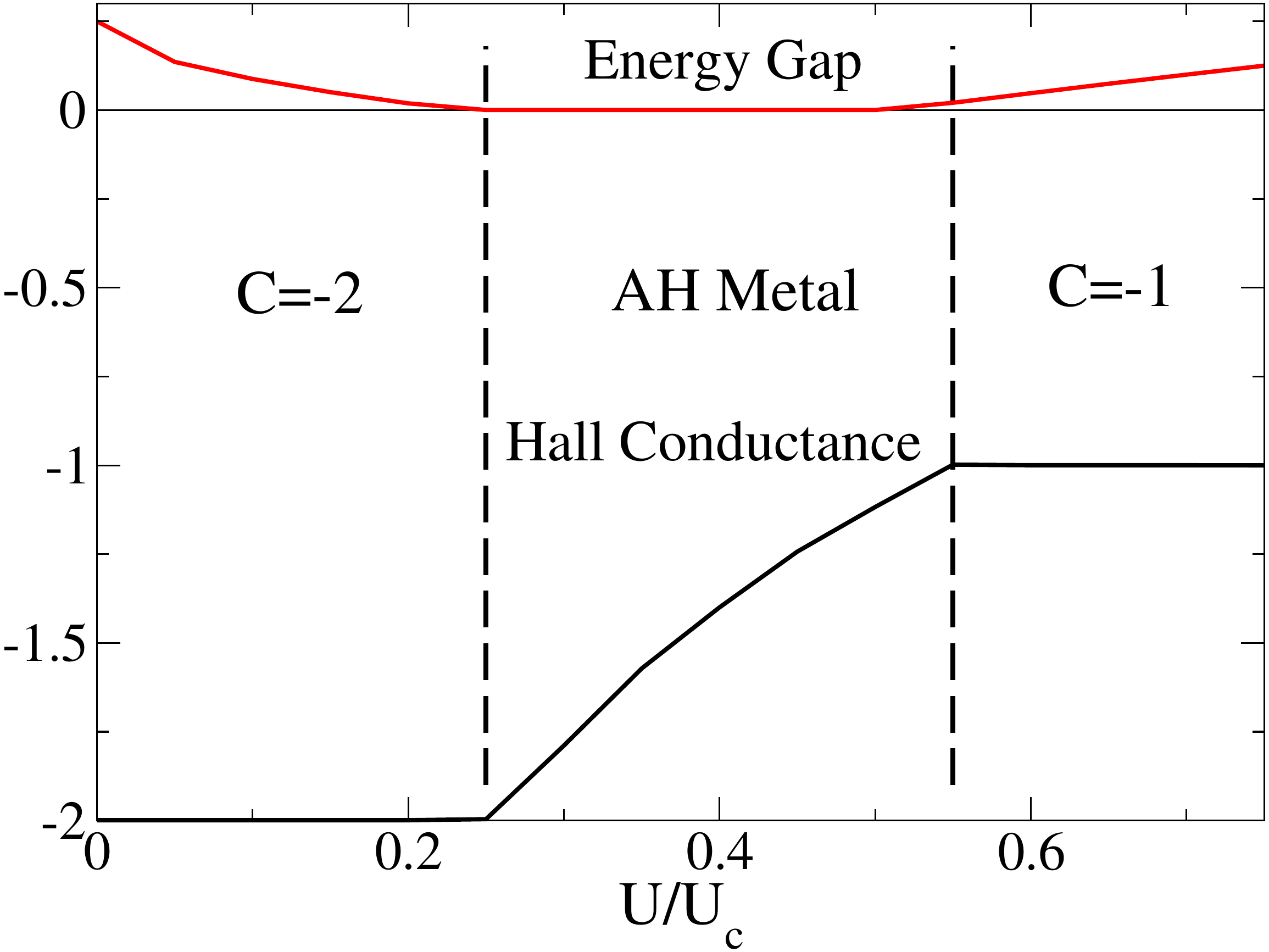}
\end{centering}
\caption{(color online). Topological transition between two topologically different
regimes through an anomalous Hall (metallic) phase as a function of $U/U_c$.
The  parameters are: $t=1, t_1=1.5$, $t_2=1.6$, $\delta=0$,
$\alpha=5$. For these parameters $U_c=12.6 t$. The Hall conductance is expressed
in units of $e^2/h$. The energy gap is the indirect gap between the two bands.
}
 \label{graphHall}
\end{figure}

Some of the models for flat bands that have been proposed in the literature
with the purpose of realizing the 
fractional anomalous quantum Hall effect,
have topological numbers similarly unstable 
with respect to the Hubbard interaction  before the Mott phase is attained.
For instance, the three-band model presented in Ref.~\cite{dasSarma1} and which
is expected to be realizable in an
 optical lattice \cite{sun2}  has a transition from $C=1$ to $C=0$ at   $\bar U/U_c=0.4$,
according to the above mechanism. 

{\it A model with spin and  $\mathbb{Z}$  topological number.---}We 
consider a system with a spinful orbital at each site described by a Hamiltonian
of the form of Eq.~(\ref{H}) with the matrices $\boldsymbol \tau$ replaced by 
the matrices acting on spin space,
$\boldsymbol \sigma$, and with
\begin{eqnarray}
h_x &=&  \alpha \sin k_y   \,, \qquad
h_y = -\alpha \sin k_x  \,, \nonumber\\
h_z &=& 4 t_2\cos k_x \cos k_y
    +2 t_1 \left( \cos k_x +  \cos k_y \right)  + \delta \,,
\label{model3}
\end{eqnarray} 
where $h_0(\boldsymbol k)=-2 t\left[\cos(k_x)+\cos(k_y)\right]$. 
Here $\alpha$ is a Rashba spin-orbit coupling,  the term proportional to
$t_1$ may be seen as an intrinsic spin-orbit coupling and $\delta$ as a
uniform magnetic field.
Time reversal implies in this case
$h_i(\boldsymbol k) \rightarrow -h_i(-\boldsymbol k)$. The term $h_{z}$ in Eq.~(\ref{model3}) breaks TRS. 
In general, due to the presence of the $h_0(\boldsymbol k)$
term there is an indirect band overlap and the system is metallic. Even though the bands are
topologically non-trivial they will be 
 in general partially filled. Considering 
$h_0(\boldsymbol k)=0$ and placing the chemical potential at zero energy one naturally
gets a half-filled band insulator with non-zero Chern number.
The  expression in Eq.~(\ref{chern}) gives
$C= 2$ if $| t _1| < |t_2 + \delta/4 |$; 
the Chern number reduces to   $C= 1$
if  $| t_1 | > |t_2 + \delta/4 |$. Turning on the Hubbard interaction, a transition
 between the different topological phases may be obtained as in the
previous model at a  suitable $\bar U<U_c$.
Considering $h_0$, one possible way to find an insulating phase is by having
a large Rashba term. By varying $t_2$ one finds regimes 
where the gap becomes zero and a band overlap occurs, hence a metallic state. 
Interestingly, increasing further $t_2$
takes the system back to an insulating phase with different Chern number. The width of this
metallic phase depends on the Rashba coupling. Starting from a $C=2$ phase and turning
on the Hubbard interaction we induce a sequence of transitions from the topological insulator
through an anomalous Hall metal and back to another insulating phase with different Chern
number. This is illustrated in Fig.~\ref{graphHall} where we plot the indirect gap between the two
bands and the Hall conductance as a function of the Hubbard coupling. There is a smooth crossover
between the two quantized values of the Hall conductance in the two insulating phases. The Hall
conductance is evaluated 
using a Kubo formula.

\begin{figure}
\begin{centering}
\includegraphics[width=7cm]{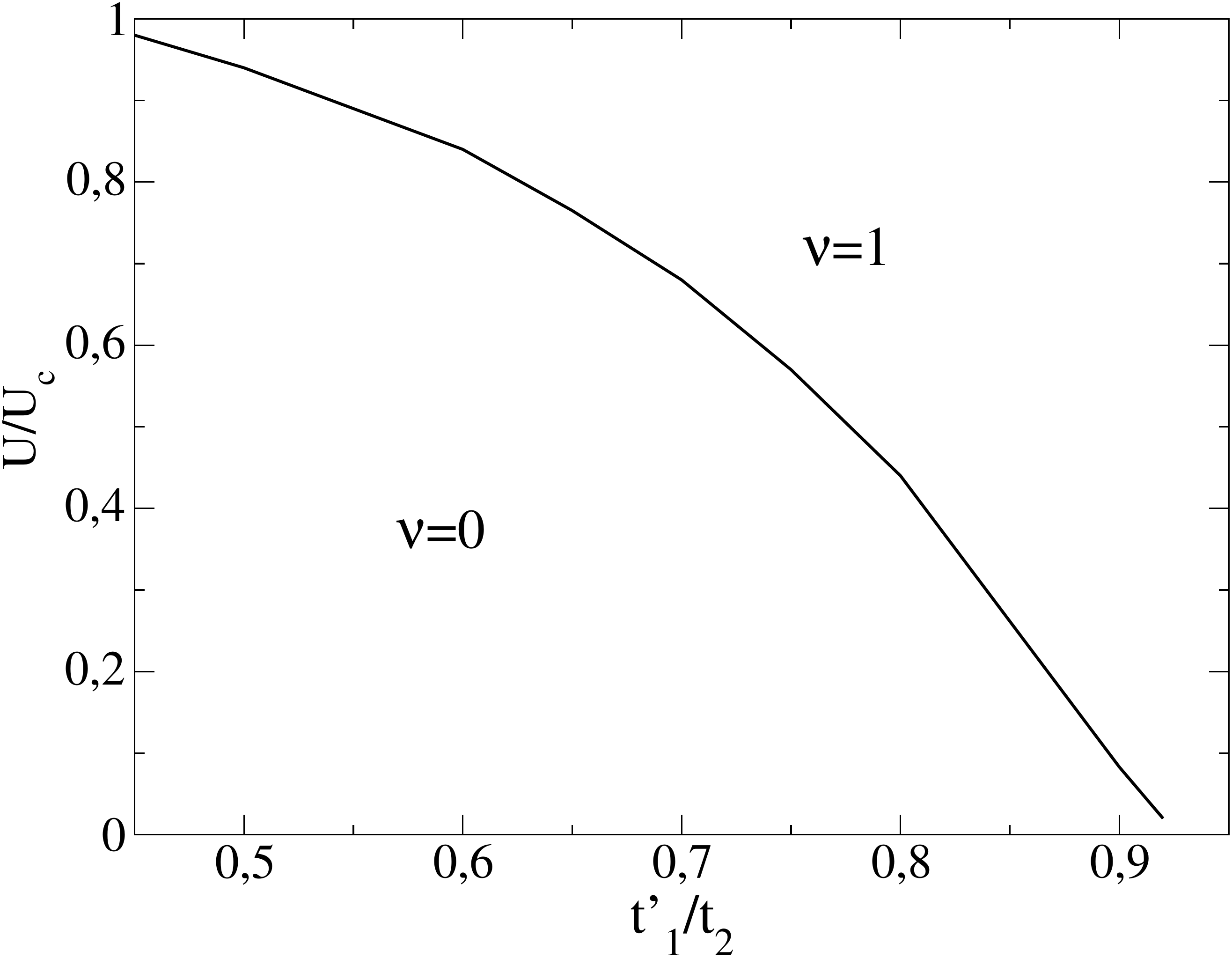}
\end{centering}
\caption{(color online). Topological phase diagram    
for the model given in Eq.~(\ref{model2}) with a Hubbard interaction. The  parameters are: $t_1=1$, $t_2=0.7$, $\delta=0$,
$t_\perp=-1$. In the phase $\nu=0$ the spin Hall conductance   $\sigma_{xy}^s = e/\pi$
whereas   $\sigma_{xy}^s = e/(2\pi)$ in the phase $\nu=1$.}
 \label{graphZ2}
\end{figure}

{\it Model with $\mathbb{Z}_2$ topological number.---}We 
extend the  model in Eq.~(\ref{model1} ) so as to describe 
 a bilayer square lattice, with one orbital per site and  
where the superscripts in  $\tau^{ss'}$ denote  layer indices. 
We introduce spin by coupling the
 spin operator $\sigma_z$ to the terms that break TRS.  
The model reads
\begin{eqnarray}
h_x &=&
   \sqrt{2} t_1  \left( \cos k_x + \cos k_y \right) + t_\perp\,,
\nonumber\\
h_y 
&=&  \sqrt{2} t_1  \left( \cos k_x - \cos k_y \right)  \sigma_z  \,,  \label{model2}\\
h_z &=& 4t_2 \sin k_x \sin k_y + 2 t_1' \left(  \sin k_x + \sin k_y  \right) \sigma_z + \delta
\nonumber
 \,.
\end{eqnarray} where $t_\perp$ 
is real and 
couples the atom in lattice site $j$ of one layer (denoted by an index $s$) 
to the one that  
sits directly above it in the other layer $s'\neq s$.  
Although TRS has been restored by 
the spin-orbit coupling, the model still lacks particle-hole symmetry, however.
This puts the model in the AII class \cite{ludwig}.   
Studies of the Mott insulating phase for systems with spin-orbit coupling and TRS, using other techniques, exist in the literature \cite{leHur}. 
  The half filled system  is  a 
  $\mathbb{Z}_2$ topological insulator
 characterized by a  topological number $\nu$ which is 
 given by the parity of the  Chern number: $\nu =0$ if $C=2$  and  $\nu =1$ if $C=1$.  
Equation~(\ref{hfx}) for this case reads
\begin{eqnarray}
 H(\bar f X,  f X^*) = \sum_{ij}\sum_{ss'}
{\boldsymbol h}_{ij}\cdot {\boldsymbol \tau}^{ss', \sigma}  \bar f_{i,s,\sigma}  f_{j,s',\sigma}  X_{is}X_{js' }^*\,,\nonumber
\label{hfxbilayer}
\end{eqnarray}
where $\sigma=\uparrow,\downarrow$ denotes the spin projections. We make a similar decoupling 
of the X-bosons and fermions as above, with the difference that the bosons now have an extra layer
index, $X_{is}$. 
A calculation with model parameters:
$t_1=1$, $t_2=0.7$, $t_1'=0.56$, $t_\perp=-1$ yields a transition from $\nu=0$ to $\nu=1$
for $\bar U=0.44 U_c$ with 
$U_c=4.3$. If  $t_1'=0.42$ then one obtains $\bar U=0.84 U_c$ with 
$U_c=4.2$.  
In Fig.~\ref{graphZ2} we show the phase diagram so obtained. 
Because of TRS, no Hall conductivity exists, but the quantized spin Hall conductance,
which is given by $\sigma_{xy}^s = eC/(2\pi)$, displays
an abrupt change at the transition, since the number of edge modes  changes discontinuously
(see Fig.~\ref{fig:edges}).
Since $\nu = 0$ is not protected by additional terms with TRS then, if disorder is present, the system may actually go from zero spin-Hall conductivity to finite (quantized) spin-Hall conductivity.

It is interesting to see that in the Kane-Mele-Hubbard model, when either a trivial Dirac mass (staggered potential) \cite{haldane1988} or a nearest-neighbor Rashba spin orbit coupling \cite{KaneMele2} are present,
 the present mechanism would destroy the topological insulating phase for a certain $\bar U < U_c$. 
 This is so because, similarly to the model in Eq.~(\ref{model2}), 
 the topological insulator in the Kane-Mele model is induced by a second nearest neigbor hopping, 
 which is thus susceptible to be renormalized by the interaction. 
 In graphene, this hopping mimics an intrinsic spin-orbit coupling \cite{KaneMele2, paco} 
 which is already very small. 
 However, this mechanism might be relevant for the recently discovered silicene \cite{siliceneExp}, 
 where the intrinsic spin-orbit coupling is expected to be sizable \cite{siliceneEzawa}.

{\it Conclusions.---}We have shown that a short ranged repulsive interaction 
can induce 
a change from a system's non-trivial topological phase to another, and
presented several examples.  The effect occurs because
the renormalization of  parameters in the Hamiltonian is such that
the longer ranged hopping is decreased with respect to the short ranged.
It is an  interesting direction for  future investigations  to explore how the renormalization
group \cite{shankar} technique for interacting fermions will renormalize   
interaction and kinetic energy
parameters so as to induce changes in the topological class of a fermionic system. 

We would like to thank Pedro Ribeiro for discussions.
We acknowledge the hospitality of CSRC, Beijing, China,
where the final stage of this work has been carried out.


\end{document}